\documentstyle[aps,prl]{revtex}
\begin{document}
\bibliographystyle{unsrt}

\title{On some upper bounds for spin velocity and instability of  gapless
state of 1-d Heisenberg chains}
\author{Konstantin Kladko}
\address{
Department of Physics, Stanford University, Stanford, CA 94305
}
\date{\today}
\maketitle
\begin{abstract}
We derive upper bounds for spin velocity in half-integer-spin Heisenberg 
antiferromagnetic chains.
We relate these upper bounds to the instability of the gapless state,
which is observed in frustrated systems.

\end{abstract}

\pacs{PACS:} 

\vskip1pc



During the last decade low dimensional strongly correlated electronic
systems have become a subject of intensive experimental and theoretical
research (see \cite{fulde} and references therein).  An important class of such systems are
one-dimensional electronic chains. It is believed, that systems with an odd number of
 electrons per unit cell (and, therefore, half-integer spin per unit cell) can be in a gapless
phase.  Such a phase should have spin-1/2 fermionic excitations - spinons,
with a linear dispersion law for small values of the wave vector $k$, $E=ck$. 
In the integrable case of the spin-1/2 nearest-neighbor Heisenberg model
existence of such excitations was established by Faddeev and Takhtajan in the paper 
\cite{faddeev} . In the nonintegrable case no proof of the above picture exists, but its
correctness seems plausible. 

In this paper we consider one-dimensional antiferromagnetic  Heisenberg models with
a half-integer spin per unit cell.  Taking for granted the picture of 
gapless excitations with a linear dispersion law for small $k$, $E=ck$,
we derive an upper bound on the spin velocity $c$. It is then shown, 
that the condition $c>0$  puts a limit on the stability
of the gapless phase in the case, when the system is frustrated.  
Finally, it is  shown on examples, that the derived upper bound is rather
good,  and in the spin-1/2 case works much better, than the linear spin-wave
approximation. 

Consider  a one-dimensional half-integer-spin Heisenberg model, which
is defined on a ring. The Hamiltonian of the model reads
\begin{equation}
H= \sum_{n,l} J_{l} ({\bf S}_n \cdot {\bf S}_{n+l}).
\label{1}
\end{equation}
Below we argue that the following upper bound holds for the spin velocity of spinons $c$.
\begin{equation}
c \le - \sum_l (4/3) l^2 J_l \pi \langle {\bf S}_n\cdot {\bf S}_{n+l} \rangle
\label{ineq}
\end{equation}
Here $\langle {\bf S}_n \cdot {\bf S}_{n+l} \rangle$ denotes the spin-spin correlation function.
Since we use periodic boundary conditions, this function does not depend on the index
$n$. We note, that for the nearest-neighbor model the nonequality (\ref{ineq})   
relates the spin velocity to the energy per unit cell.    

Here we  describe a sketch of the derivation for the case of the nearest-neighbor
$S=1/2$ Heisenberg model. The full derivation is given at the end of the paper.  

1.Consider a periodic ring of $2N+1$ sites. The full spin of the ground state
$\mid \psi_0^{(N)} \rangle$ is
1/2, which
corresponds to having one spinon in the $k=0$ state. Let us denote the energy
of the ring as $E_0^{(N)}$.
 
2.We apply Lieb-Schultz-Mattis transformation \cite{mattis} to this state, obtaining a new quantum
state $\mid \psi_1^{(N)}\rangle$. We then check, that this state has a wave vector $k=\pi(1+1/(2N+1))$. 
Using this state as a trial state, and evaluating its energy, we show that in the thermodynamic
limit the energy difference between the lowest energy
state at $k=\pi(1+1/(2N+1))$ and the ground state of the system is less or equal
to 
\begin{equation}
E_1 - E_0 \le  - (4/3)  J  \pi^2 \langle {\bf S}_n\cdot {\bf S}_{n+1} \rangle/(2N+1).
\label{endif}
\end{equation}
3.Since the lowest energy spectrum $E(k)$ is $\pi$ periodic, as was proven
in \cite{kladko}, one can evaluate the spin velocity as $\left[ E(\pi+\Delta k)-E(\pi)\right]/\Delta k $,
where $\Delta k$ is $\pi/(2N+1)$. Then, using the fact, that $E(\pi) \ge E_0$ ($E_0$ is the
ground state energy) and  (\ref{endif}),
one obtains (\ref{ineq}).

Let us now study the implications of the inequality (\ref{ineq}).  
For the nearest-neighbor model the inequality (\ref{ineq}) gives $c \le - (4/3) \pi E_{per.site}\simeq 0.591 \pi J$.
The exact value of $c$ given by the Bethe Ansatz  is $0.5 \pi J$. 
Let us note that the linear spin-wave theory
gives $c \simeq 0.318 \pi J$.  This shows, that, for spin-1/2 models, the wave
function of spinons
is much closer to the modes, described above, than to the linear spin waves. We also
note, that the inequality (\ref{ineq}) may be improved, if one finds a way to modify
the LSM transformation in such a way that it does not break the rotation group
symmetry.  Studies in this direction will be reported elsewhere \cite{kladko2}

Consider now  a frustrated $S=1/2$ chain with antiferromagnetic exchanges
$J_1, J_2$. Then since the spin-velocity should be positive, the gapless
state is necessarily unstable for 
$J_1 \langle{\bf S}_n\cdot {\bf S}_{n+1}\rangle + 4 J_2 \langle {\bf S}_n\cdot {\bf S}_{n+1}\rangle  > 0$,
i.e., for 
\begin{equation}
J_2/J_1 > -0.25 \langle {\bf S}_n\cdot {\bf S}_{n+1}\rangle/\langle {\bf S}_n\cdot {\bf S}_{n+2}\rangle.   
\end{equation}
Using correlation functions ratio from the unfrustrated system we 
get a rough estimate of the critical value of $J_2/J_1$ at 0.5.
Numerical calculations \cite{white} 
put this value at about 0.4. At the transition point the
instability  leads to an appearance of a dimerized phase with a gap.
Let us now consider the large-S limit. In the limit of large S the
groundstate is the Neel state and then we get the condition
$J_2/J_1 > 0.25$. Let us note that $J_2=1/4$ is  the point, where  
the classical Neel state becomes unstable with respect to spiral order,
so the classical limit is recovered exactly. 

{\bf Derivation}

Here we use notations similar to \cite{kladko}. We first give a derivation
for the case of the nearest-neighbor $S=1/2$ model, and then show how
it is extended to the general Hamiltonian (\ref{1}).

1.We assume that the system is in the gapless antiferromagnetic phase, 
and that the ground state of a ring having $2N+1$ sites has the full spin
$S_{\textrm{full}}=1/2$. 
We denote the ground state of the ring as $\mid \psi^{(N)}_0 \rangle$. We choose 
the coordinate system in such a way, that the $z$-component of the
full spin $S^z_{\textrm{full}}$ is equal to $1/2$.   

2.Let us write the state $\mid \psi^{(N)}_0 \rangle$ as a linear combination of $S_{\textrm{full}}^z=1/2$ spin configurations
$\mid \psi^{(N)}_0 \rangle = \sum_\sigma A_\sigma \mid \sigma \rangle$.
Any $S_{\textrm{full}}^z=1/2$  configuration $\mid \sigma \rangle$
is characterized by  $N+1$ numbers $x_1 < x_2 < x_3 ...$ showing the positions of fictitious particles, each of them increasing  $S^z$ on a given site by one.
The vacuum of the system is assumed to have all spins down. Since the system is periodic, the $x$'s are defined only $mod$ $2N+1$.
Consider the state 
\begin{equation}
\label{4}
\mid \psi^{(N)}_{1}> = \sum_\sigma e^{\frac{2\pi i}{2N+1}\left( x_1+x_2+...\right) }  A_\sigma \mid \sigma \rangle . 
\end{equation}
Adding $2N+1$ to any of the $x$'s does not change  the  $\mid \psi^{(N)}_{1}>$. Therefore, 
the mentioned freedom of defining the $x$'s is satisfied.
If  the translation operator operator $T:n\rightarrow n+1$ acts on the state  $\mid \psi^{(N)}_{1}> $, then all $x$'s are incremented by 1. Therefore, an
additional phase factor $e^{i\pi\left[1+ 1/(2N+1)\right]}$ is acquired, and the state $\mid \psi^{(N)}_{1}> $ has a wave 
vector $\pi\left[1+ 1/(2N+1)\right]$.

Let us evaluate the expectation value of energy $\langle \psi^{(N)}_{1} \mid H \mid  \psi^{(N)}_{1} \rangle $.
The Ising part of the Heisenberg Hamiltonian does not flip spins.  In contrast the
exchange part may change the position of one spin-up fictitious particle, which  amounts to a phase factor
$e^{\pm \frac{2\pi i}{2N+1}}$, depending on a direction of the move.  This explains the result
of a  formal calculation, which gives: 
\begin{eqnarray}
\label{6}
\langle \psi^{(N)}_{1}  \mid H \mid  \psi^{(N)}_{1}  \rangle  = \\ 
\nonumber
= \sum_n J \left[ 1/2 \left(  e^{\frac{2\pi i}{2N+1}}  \langle \psi^{(N)}_{0} \mid  S^+_n S^-_{n+1} \mid  \psi^{(N)}_{0}  \rangle  +
  e^{-\frac{2\pi i}{2N+1}}  \langle \psi^{(N)}_{0}  \mid  S^+_{n+1} S^-_n  
\mid  \psi^{(N)}_{0}  \rangle  \right)  +
  \langle \psi^{(N)}_{0}  \mid  S^z_{n+1} S^z_n \mid  \psi^{(N)}_{0}  \rangle  \right].
\end{eqnarray}
Taking a thermodynamic limit of (\ref{6}) one finds
\begin{eqnarray}
\label{9}
\langle \psi^{(N)}_{1}  \mid H \mid  \psi^{(N)}_{1}  \rangle  - \langle \psi^{(N)}_{0}  \mid H \mid  \psi^{(N)}_{0}  \rangle  =
\\ \nonumber 
=(2N+1) \frac{J}{2} ( e^{- \frac{2\pi i}{2N+1}}  \langle \psi^{(N)}_{0}  \mid  S^+_{n+1} S^-_n \mid  \psi^{(N)}_{0}  \rangle  +
e^{\frac{2\pi i}{2N+1}}  \langle \psi^{(N)}_{0}  \mid  S^+_n S^-_{n+1} \mid  \psi^{(N)}_{0}  \rangle - \\   
\nonumber
-  \langle \psi^{(N)}_{0}  \mid  S^+_{n+1} S^-_n \mid  \psi^{(N)}_{0}  \rangle   
- \langle \psi^{(N)}_{0}  \mid  S^+_n S^-_{n+1} \mid  \psi^{(N)}_{0}  \rangle )
= 
i J \pi  \langle \psi^{(N)}_{0}  \mid  S^+_n S^-_{n+1} \mid  \psi^{(N)}_{0}  \rangle 
 - \langle \psi^{(N)}_{0}  \mid  S^+_{n+1} S^-_n \mid  \psi^{(N)}_{0}  \rangle ) 
 - \\ \nonumber - \frac{\pi^2}{2N+1}
 \left( \langle \psi^{(N)}_{0}  \mid  S^+_n S^-_{n+1} \mid  \psi^{(N)}_{0}  \rangle 
 + \langle \psi^{(N)}_{0}  \mid  S^+_{n+1} S^-_n \mid  \psi^{(N)}_{0}  \rangle \right)   +
O(1/(2N+1)^2)
\end{eqnarray}
The operator $S^+_n S^-_{n+1} - S^+_{n+1} S^-_n$ is proportional to the current operator
of the $z$-component of the spin \cite{kladko}. Since the ring 
has reflection symmetry, one can always choose a ground state in such a way, that the 
expectation value of this operator is exactly zero. Then one has 
\begin{eqnarray}
\label{10}
\langle \psi^{(N)}_{1}  \mid H \mid  \psi^{(N)}_{1}  \rangle 
 - \langle \psi^{(N)}_{0}  \mid H \mid  \psi^{(N)}_{0}  \rangle  = \\
\nonumber
= -\frac{\pi^2}{2N+1}
 \left( \langle \psi^{(N)}_{0}  \mid  S^+_n S^-_{n+1} \mid  \psi^{(N)}_{0}  \rangle 
 + \langle \psi^{(N)}_{0}  \mid  S^+_{n+1} S^-_n \mid  \psi^{(N)}_{0}  \rangle \right)   +
O(1/(2N+1)^2)= 
\\
\nonumber 
= - (4/3)  J  \pi^2 \langle {\bf S}_n\cdot {\bf S}_{n+1} \rangle/(2N+1)+
O(1/(2N+1)^2).
\end{eqnarray}

3.Now we note, that, as was proven in \cite{kladko}, the spectrum of the lowest energy
excitations $E(k)$ is $\pi$-periodic in the thermodynamic limit. Therefore, one can find the spin-velocity using
the formula $c=\left[ E(\pi + \Delta k) - E(\pi) \right]/ \Delta k$, where $\Delta k$ 
is $\Delta k = \pi/(2N+1)$. The energy of the lowest energy state at $k=\pi+ \Delta k$ 
is less or equal to the expectation value of energy for the trial state 
$\mid  \psi^{(N)}_{1}  \rangle $. The energy of the lowest energy state at $k=\pi$ is more
or equal to the energy of the ground state (equality is reached in the thermodynamic limit).
Based on these considerations, taking (\ref{10}) and dividing it by $\Delta k$ we obtain in the thermodynamic
limit  $N \rightarrow \infty$ an upper bound for the spin velocity $c$
\begin{equation}
c \le - (4/3) J  \pi \langle {\bf S}_n\cdot {\bf S}_{n+1} \rangle
\label{ineq2}
\end{equation}
Now for the general case (\ref{1}), repeating calculations above and introducing a sum
over the index $l$ in the Hamiltonian, one arrives at the general inequality (\ref{ineq}). 
One can also repeat the calculations for the case of a general half-integer spin $s$,
again obtaining (\ref{ineq}). Let us note that for large  $S$ the upper bound (\ref{ineq})
scales as $S^2$, whether the linear-spin wave approximation, which becomes exact in
large-S limit, gives the spin velocity, which scales like $S$, in particular for the
nearest-neighbor model one has $c=2SJ$. For the $S=3/2$ nearest-neighbor model 
numerics give $c=3.87 J$, linear spin waves give $c=3J$, and the upper bound
(\ref{ineq}) is  $11.8 J$ (we use the exact value of the ground state energy from 
\cite{hallberg}).  This shows, that the upper bound derived here is interesting
only for $S=1/2$ chains.

To conclude, we have derived an upper bound for the spin velocity of half-integer-spin
Heisenberg antiferromagnetic chains and have shown its relation to the instability
of the gapless state in the presence of frustration.

\end{document}